\documentstyle[tighten,aps,epsfig,doublespace,preprint,eqsecnum,epsf]{revtex}

\begin{document}
\draft
\title{Phase transition and spin-wave dispersion in
quantum Hall bilayers at filling factor $\nu=1$}
\author{Anton Burkov$^{a,b}$,
John Schliemann$^{a,b}$\thanks{e-mail: joschlie@physics.utexas.edu},
A.~H. MacDonald$^{a,b}$,
and S.~M. Girvin$^{b}$}
\address{
$^{a}$Department of Physics, The University of Texas, Austin, TX 78712\\
$^{b}$Department of Physics, Indiana University, 
Bloomington, IN 47405}
\maketitle
\begin{abstract}
We present an effective Hamiltonian for a bilayer
quantum Hall system at filling factor $\nu=1$ neglecting charge fluctuations.
Our model is formulated in terms of spin and pseudospin operators
and is an exact representation of the system within the above approximation.
We analyze its low-lying excitations in terms of spin-wave theory. Moreover
we add to previous first-principle exact-diagonalization studies 
concentrating on the quantum phase transition seen 
in this system.
\end{abstract}

\section{Introduction}
At small layer separations, the ground state of a $\nu=1$ bilayer
quantum Hall system exhibits spontaneous interlayer phase coherence. 
Interest in this issue has been renewed by
intriguing tunneling transport measurements by Spielman {\it et al.}
showing a very pronounced conductance peak at zero bias voltage
\cite{eisen00} for sufficiently small values of the ratio of layer separation
to magnetic length.

The quantum phase transiton underlying this phenomenon has been 
investigated recently in an exact diagonalization study using the
spherical geometry \cite{schlie01}. The results suggest that a single phase 
transition, likely of first order, 
separates incompressible states with strong interlayer correlations 
from compressible states with weak interlayer correlations.

In the present work we report on a study of $\nu=1$ quantum Hall bilayer
using a different approach introducing an 
effective spin-pseudospin model on an imaginary
lattice in the lowest Landau level (LLL) orbital space (von Neumann lattice).
We also add further exact diagonalization results concerning the position
and order of the compressible-incompressible transition.

\section{Effective spin-pseudospin-model}
A single-particle state of electron in a $\nu=1$ bilayer is specified 
by three quantum numbers: lowest Landau level (LLL) orbit-center quantum
number $i$, spin $\sigma$ and pseudospin $\tau$, describing the 
layer degree of freedom. 
However, in the incompressible state bilayer has a gap for charged excitations 
and therefore one can assume that only the spin and pseudospin degrees of 
freedom are relevant for the low temperature physics.
The microscopic Hamiltonian reads 
${H}={H}_{{1\rm P}}+{H}_{\rm Coul}$, 
where ${H}_{{\rm Coul}}=V^{S}+V^{D}+$ 
represents the usual Coulomb interaction within ($V^{S}$) 
and between layers ($V^{D}$), and the single-particle Hamiltonian 
${H}_{{1\rm P}}$ implements tunneling of electrons between the layers
with an amplitude $\Delta_{t}$ and couples electron spins to the 
perpendicular magnetic field with amplitude $\Delta_z$.  
One would like to eliminate the irrelevant charge degree of freedom
from the microscopic Hamiltonian and arrive at an effective model containing 
only spin and pseudospin variables. 
The most convenient and mathematically rigorous way to do this is provided 
by the functional integral approach. 
One notices that the no-charge-fluctuations subspace of the system is spanned 
by single Slater determinant many body wave functions of the form
$|\Psi[z]\rangle = \prod_i \left(\sum_{k=1}^4 z_{ik} c^{\dagger}_{ik}
\right)| 0 \rangle$.
Here $i$ is the LLL orbital quantum number and $k$ is a 4-component 
spinor index describing the combined spin-pseudospin degree of freedom. 
$k=1$ means electron in the top layer with an up spin,
$k=2$---top layer down spin, $k=3$---bottom layer up spin, $k=4$---bottom 
layer down spin.
$c^{\dagger}_{ik}$ is the creation operator for an electron in the 
LLL orbital state $i$ and with 4-spinor index $k$. 
$z_{ik}$ are complex amplitudes, satisfying the normalization condition
$\sum_{k=1}^4 |z_{ik}|^2=1$

One can therefore formally write the partition function of the system as
a functional integral over this overcomplete set of Slater determinants.
After considerable algebra one arrives at the
following effective Hamiltonian
\begin{eqnarray}
\label{6}
H&=&-\sum_i\left[\Delta_t T^x_i+\Delta_z S^z_i\right]
\nonumber\\
&+&\sum_{ij} \Big[ (2H_{ij}-\frac{1}{2}F^S_{ij})T^z_iT^z_j-
\frac{1}{2}F^D_{ij}{\bf T}^{\perp}_i {\bf T}^{\perp}_j\nonumber\\
&-&\frac{1}{2}F^S_{ij}{\bf S}_i {\bf S}_j
2 F^S_{ij}({\bf S}_i {\bf S}_j) T^z_i T^z_j\nonumber\\
&-&2F^D_{ij}({\bf S}_i {\bf S}_j)({\bf T}^{\perp}_i {\bf T}^{\perp}_j)\Big]
\end{eqnarray}
Here $F^{S,D}_{ij}=F^{+}_{ij}\pm F^{-}_{ij}$,
and $H_{ij}=\langle ij|V_-|ij\rangle, F^{\pm}_{ij}=\langle ij|V_{\pm}|ji
\rangle$ are the direct and exchange matrix elements of the Coulomb 
interaction $V^{\pm}=(V^{S}\pm V^{D})/2$, 
and ${\bf S}_{i}$ and ${\bf T}_{i}$ are local spin and 
pseudospin operators, respectively.

Our spin-pseudospin model is defined on a ``lattice'' with ``sites'' labeled 
by the LLL orbital quantum numbers $i$. 
So far we have not specified the orbital basis we are using.
It is obvious that the usually used orbit-center quantum numbers in Landau or
symmetric gauges are not good choices here.
Both of these basis sets introduce an artificial gauge-dependent asymmetry 
into the problem.
One would like to use a basis more appropriate for a spin model: that 
of Wannier-like functions localized at the sites of imaginary square 
lattice (von Neumann lattice) with lattice constant $\sqrt{2\pi l^2}$ to 
accomodate exactly
one electron or corresponding spin-pseudospin pair per site at filling factor
$\nu=1$.
It is not obvious in advance that such a basis exists.
Strong magnetic field imposes certain restrictions on the localization 
properties of magnetic orbitals~\cite{Zak,Thouless}, 
and it is well established for example
that a set of linearly independent {\it and} exponentially localized 
single-particle orbitals in the LLL does not exist.  
However, it turns out to be possible~\cite{Ferrari,Efros} to construct
a complete orthonormal set of Wannier-like eigenfunctions, which,
although not exponentially localized, have a well defined Gaussian core
with power law falloff at large distances. 
Following~\cite{Ferrari,Efros} we will call them magnetic Wannier functions. 
The procedure one uses to construct such a basis set is very much like
the one used to construct the usual Wannier functions in a crystal. 
One starts from the set of minimum uncertainty wavepackets 
for electrons in the LLL, centered at the sites of the square lattice
described above. 
The difference from the case of a crystal here is that this set 
is {\it overcomplete}, as was shown by Perelomov~\cite{Perelomov}.
One then constructs Bloch functions from linear combinations of the 
minimum uncertainty wavepackets and Fourier transforms them to obtain 
the Wannier functions. 
There are subtleties in this procedure and 
we refer the reader to the original papers~\cite{Ferrari,Efros} for further 
details.

In this paper we will present only the simplest linear spin wave
results for our spin-pseudospin model, which are applicable at zero 
temperature.  
A detailed study of finite temperature properties of (\ref{6}) 
will be presented in a forthcoming publication~\cite{Burkov01}.
As evident by direct inspection, the interaction part of (6) has the correct 
$SU(2)_{spin}\times U(1)_{pseudospin}$ symmetry. 
Correspondingly there are two Goldstone modes associated with the 
spontaneous breaking of this symmetry. 
Their dispersions are given by
\begin{eqnarray}
\label{7}
E_k^{spin}&=&\Delta_z+F^+_0-F^+_k\nonumber\\
E_k^{pseudospin}&=&\Big((\Delta_t+F^D_0+H_k-F^+_k)^2\nonumber\\
& &-(H_k-F^-_k)^2\Big)^{\frac{1}{2}}
\end{eqnarray}  
In Fig.1 we show dispersions (\ref{7}) evaluated for a 20$\times$20 square 
lattice using realistic values for the tunneling and Zeeman splittings
at interlayer separation $d=1.4l_{B}$ with $l_{B}$ being the magnetic length.
One can see that the spin dispersion is quadratic and the pseudospin one
is linear at small values of the wavevector in accordance with broken 
$SU(2) \times U(1)$ symmetry.     
Another thing to notice is that the pseudospin gap is appreciably 
larger than the spin gap even though the bare values of tunneling and 
Zeeman coupling are the same. 
This makes invalid (at finite temperature) the usual argument about spin 
fluctuations being frozen out by the magnetic field \cite{halperin83}.
The dip in the pseudospin mode dispersion at the Brillouin zone boundary 
signals softening of the pseudospin mode due to the development of 
antiferromagnetic instability,
which at large enough interlayer separations ($\sim 1.45$ in our model)
leads to the transition to the compressible state.
  
\section{Results on the phase transition}
Our exact diagonalization results are obtained for a bilayer quantum Hall
system in the spherical geometry containing spin-polarized electrons.
We will consider both the case of zero and finite 
width $w$ of the two quantum wells \cite{schlie01}.
In the exact diagonalization data a transition between a 
compressible ground state with weak interlayer correlations to an 
incompressible strongly correlated phase is signalled by a maximum
of the fluctuation $\Delta T^{x}=
\sqrt{\langle{T^{x}}^{2}\rangle-\langle T^{x}\rangle\langle T^{x}\rangle}$
of the ground state pseudospin magnetisation
along with its susceptibility $\chi=d\langle T^{x}\rangle/d\Delta_{t}$
\cite{schlie01}. These maxima grow rapidly with increasing system indicating
a quantum phase transition. The positions of these maxima as a function of
tunneling gap and layer separation (measured in units of the magnetic length)
define finite-size phase boundaries (cf. figure 3 in Ref. \cite{schlie01}).
The spectacular phenomenon found by Spielman {\it et al.} \cite{eisen00}
occured in samples
with extremely small tunneling amplitude close to the limit of vanishing 
tunneling where {\em spontaneous} interlayer phase coherence arises.
Therefore, the critical layer separation at zero tunneling is of particular
interest. The values of this quantity obtained for finite systems with up
to twelve electrons form a rapidly converging data series 
(cf figure 4 in Ref. \cite{schlie01}). The critical values $d_{c}$
for the layer
separation at zero tunneling are shown in figure \ref{diagfig1} as a
function of the ratio $w/d$ of well width to layer separation. A value
of $w/d=0.65$ corresponds to the samples used in Refs. \cite{eisen00},
where the experimental value of $d=1.83$ for the critical layer separation
agrees very well with the exact diagonalization result of $d=1.81$. 

In order to further investigate the order of the quantum phase transition,
we introduce  the ratio
\begin{equation}
\omega_{N}=\frac{2\left(\Delta T^{x}\right)^{2}_{N}}
{\left(d\langle T^{x}\rangle/d\Delta_{t}\right)_{N}}\,,
\label{omega}
\end{equation}
where the subscript $N$ refers to the system size. 
As discussed in Ref. \cite{schlie01}, 
this type of ratio defines a characteristic energy scale of the
system at the phase boundary and
should prove to be a powerful general 
tool in the analysis of any quantum phase transition.

For a continuous phase transition one would clearly expect $\omega_{N}$ to
vanish at the phase boundary for an infinite system, while a finite limit 
$\lim_{N\to\infty}\omega_{N}$ is indicative of a finite energy scale, 
i.e. a first order transition. From our finite-size data for $\omega_{N}$
(evaluated at vanishing tunneling and $d=d_{c}(N)$) 
shown in figure \ref{diagfig2} we conclude
that this quantity extrapolates for $N\to\infty$ 
to a rather substantial non-zero value 
of order $0.05 e^{2}/\epsilon l_{B}\sim 5$K for
all values of $w$ considered here. Along with the arguments and experimental
findings given so far, this result strongly suggests that the
bilayer quantum Hall system at filling factor $\nu=1$ undergoes a single
first order phase transition as a function of the ratio of layer separation
and magnetic length at all values of the tunneling 
amplitude. The phase boundary separates a phase with strong 
interlayer correlation (and a finite gap for charged excitations) from a 
phase with weak interlayer correlations and vanishing gap for charged 
excitations.

\section*{Acknowledgements}

This work was supported by the Deutsche Forschungsgemeinschaft, 
the National Science Foundation, and the Welch Foundation.

\begin{figure}
\centerline{\includegraphics[width=7.6cm]{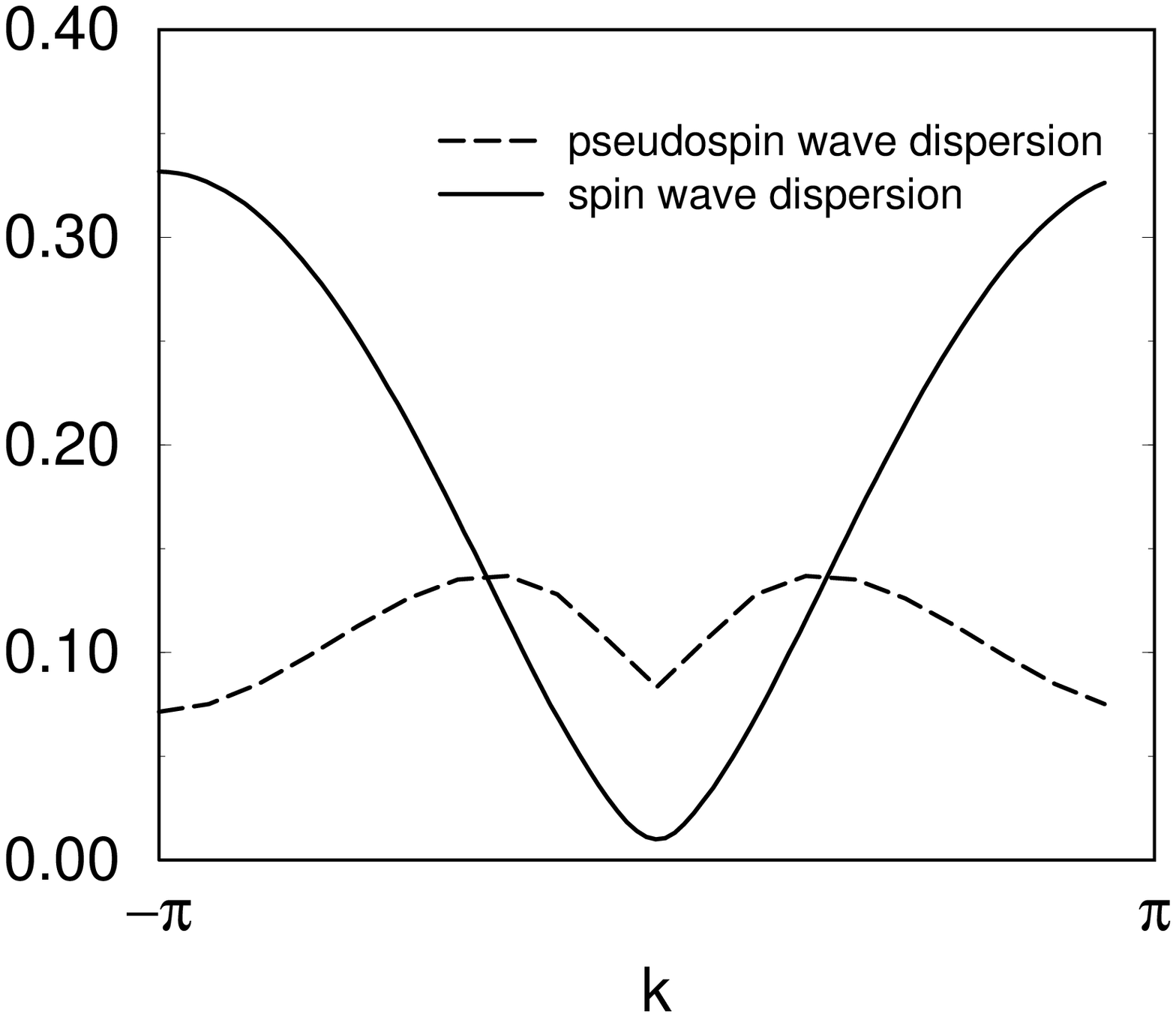}}
\caption{Spin and pseudospin mode dispersions in the (100) direction 
for $\Delta_t=\Delta_z=0.01$ in units of $e^2/\epsilon l$ and $d=1.4l_{B}$.}
\end{figure}

\begin{figure}
\centerline{\includegraphics[width=7.6cm]{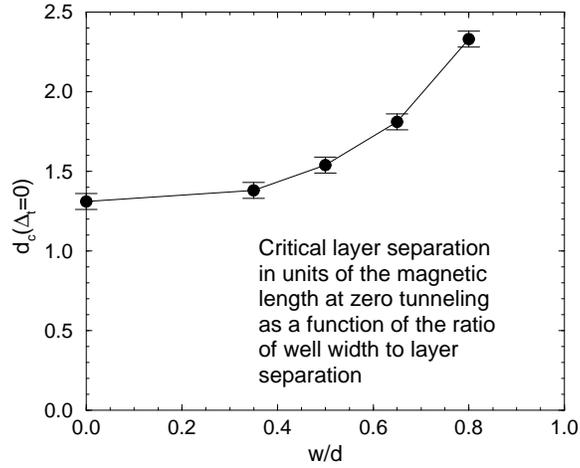}}
\caption{Critical layer separation (extrapolated in the thermodynamic
limit) at zero tunneling as a function of the ratio of well width to
layer separation.
\label{diagfig1}}
\end{figure}

\begin{figure}
\centerline{\includegraphics[width=7.6cm]{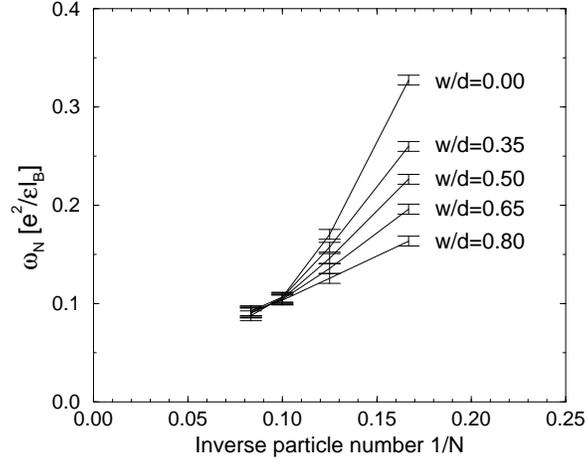}}
\caption{The averaged excitation energy $\omega_{N}$ as a function of the
system size for various ratios of well width $w$ to layer separation
$d$. Assuming that the these data curves remain of a convex shape also
for larger $N$, one concludes that $\omega_{N}$ extrapolates to
 finite values (being of order 0.05 in units of the Coulomb energy scale) 
for all values of $w/d$. 
\label{diagfig2}}
\end{figure}

\end{document}